\documentclass{emulateapj}
\usepackage{apjfonts}
\usepackage{epsfig}

\usepackage[latin1]{inputenc}
\usepackage{lscape}
\shorttitle{Multiple Radio Sources in the L723 Outflow}
\shortauthors{Carrasco-Gonz\'alez et al.}

\begin{document}

\title{A Multiple System of Radio Sources at the Core of the L723
Multipolar Outflow}

\author{Carlos~Carrasco-Gonz\'alez\altaffilmark{1}, Guillem~Anglada\altaffilmark{1}, 
Luis~F.~Rodr\'{\i}guez\altaffilmark{2}, Jos\'e~M.~Torrelles\altaffilmark{3}, 
Mayra~Osorio\altaffilmark{1} and~Jos\'e~M.~Girart\altaffilmark{4}}

\altaffiltext{1}{Instituto Astrof\'{\i}sica Andaluc\'{\i}a, CSIC, Camino
Bajo de Hu\'etor 50, E-18008 Granada, Spain; charly@iaa.es,
guillem@iaa.es, osorio@iaa.es}

\altaffiltext{2}{Centro de Radioastronom\'{\i}a y Astrof\'{\i}sica UNAM,
Apartado Postal 3-72 (Xangari), 58089 Morelia, Michoac\'an, M\'exico;
l.rodriguez@astrosmo.unam.mx}

\altaffiltext{3}{Instituto de Ciencias del Espacio (CSIC) and Institut
d'Estudis Espacials de Catalunya, Facultad de F\'{\i}sica, Planta 7a,
Universitat de Barcelona, Av. Diagonal 647, E-08028 Barcelona, Spain; 
torrelles@ieec.fcr.es}

\altaffiltext{4}{Instituto de Ciencias del Espacio (CSIC) and Institut
d'Estudis Espacials de Catalunya, Campus UAB, Facultat de Ci\`encies,
Torre C-5 parell, E-08193 Bellaterra, Spain; girart@ieec.uab.es}

\received{2006 November 17}
\accepted{2007 December 6}

\begin{abstract}

We present high angular resolution Very Large Array multi-epoch continuum
observations at 3.6 cm and 7 mm towards the core of the L723 multipolar
outflow revealing a multiple system of four radio sources suspected to be
YSOs in a region of only $\sim$4$\arcsec$ (1200 AU) in extent. The 3.6 cm
observations show that the previously detected source VLA 2 contains a
close (separation $\simeq 0\rlap.''29$ or $\sim$90 AU) radio binary, with
components (A and B) along a position angle of $\sim 150^\circ$. The
northern component (VLA 2A) of this binary system is also detected in the
7 mm observations, with a positive spectral index
 between 3.6 cm and 7 mm. In addition, the source VLA 2A is associated
with extended emission along a position angle of $\sim 115^\circ$, that we
interpret as outflowing shock-ionized gas that is exciting a system of HH
objects with the same position angle. A third, weak 3.6 cm source, VLA 2C,
that is detected also at 7 mm, is located $\sim$0$\farcs$7 northeast of
VLA 2A, and is possibly associated with the water maser emission in the
region. The 7 mm observations reveal the presence of an additional source,
VLA 2D, located $\sim$3$\farcs$5 southeast of VLA 2A, and with a 1.35 mm
counterpart. All these radio continuum sources have a positive spectral
index, compatible with them being YSOs. We also propose that the high
velocity CO emission observed in the region could be the superposition of
multiple outflows (at least three independent bipolar outflows) excited by
the YSOs located at the core, instead of the previous interpretations in
terms of only one or two outflows.

\end{abstract}

\keywords{ISM: individual (L723) --- ISM: jets and outflows --- radio 
continuum: ISM --- stars: formation}

\section{Introduction}

 L723 is an isolated dark cloud at a distance of 300 $\pm$ 150 pc (Goldsmith et al. 1984). Located in this cloud
is a Class 0 source, IRAS 19156+1906, with a luminosity of $\sim$3.4 L$_\sun$ (Dartois et al. 2005). The IRAS
source is associated with a CO outflow, first mapped by Goldsmith et al. (1984). The outflow shows a peculiar
quadrupolar morphology, consisting of two pairs of red-blue lobes with a common center, as is clearly seen in the
maps of Avery et al. (1990). The larger pair of lobes extends along a direction with a position angle (P.A.)
$\simeq$ 100$^\circ$, while the smaller pair extends along a direction with a P.A. $\simeq$ 30$^\circ$. Over the
years, the peculiar quadrupolar morphology of the L723 outflow has been interpreted as due to the presence of
either one or two driving sources (see Anglada, Rodr\'{\i}guez \& Torrelles 1996 and references therein). In the
first case, the quadrupolar morphology is attributed to limb-brightening effects in the lobes of a single bipolar
outflow, to splitting of the lobes due to interaction with ambient clumps, or to precession of the outflow axis.
In the second case, the quadrupolar structure is attributed to two independent bipolar outflows, each one driven
by a different source. 

\begin{deluxetable*}{ccccccrc}
\tabletypesize{\scriptsize}
%\rotate
\tablewidth{0pt}
%\tablenum{}
\tablecaption{Observations at 3.6 cm \label{tabla1}}
%\tablehead{}
%\tablecolumns{}
\startdata
\hline \hline 
        &              &	                   &                &   Bootstrapped                    &                    &		                        &      \\
	&	       &                           & Effective      &   Flux Density of                 & \multicolumn{2}{c}{Synthesized Beam\tablenotemark{b}} &      \\ \cline{6-7}
	& Observation  & Hour Angle Ranges         & Time on Source & Phase Calibrator\tablenotemark{a} &	  HPBW       &   \multicolumn{1}{c}{P.A.}       & rms Noise\tablenotemark{b} \\
 Epoch  & Date         &         (h)               &	  (h)       & (Jy)	                        &  (arcsec)          & \multicolumn{1}{c}{(deg)}        & ($\mu$Jy beam$^{-1}$)      \\ \hline
 1995.6 & 95-Aug-12    & $-$4.4 to $-$2.5 , 2.0 to 4.0 &   3.7      & 1.146 $\pm$ 0.006                 & 0.32 $\times$ 0.27 & $-$85                            &  11  \\
 1997.0 & 97-Jan-10    & $-$2.0 to 2.5 	           &   4.0          & 0.926 $\pm$ 0.004                 & 0.27 $\times$ 0.25 &  15                              &  11  \\
 1998.2 & 98-Mar-27    & 1.0 to 3.5, 4.3 to 5.8    &   3.7          & 0.94  $\pm$ 0.01                  & 0.34 $\times$ 0.25 &  58                              &  12  \\
 1998.4 & 98-May-26    & 1.0 to 3.5, 4.3 to 5.8    &   3.7          & 0.952 $\pm$ 0.007                 & 0.34 $\times$ 0.25 &  58                              &  11  \\
 1999.5 & 99-Jul-03    & 1.0 to 3.5, 4.3 to 5.8    &   3.7          & 1.260 $\pm$ 0.006                 & 0.39 $\times$ 0.26 &  50                              &  14  \\
 2000.9 & 00-Dec-16    & $-$5.0 to 4.5             &   8.0          & 1.65  $\pm$ 0.02                  & 0.28 $\times$ 0.27 & $-$70                            &   6  
\enddata
\tablenotetext{a}{The phase calibrator used in all the observations was 1925+211.}
\tablenotetext{b}{For naturally weighted maps.}
\end{deluxetable*}

\begin{deluxetable*}{ccccrc}[!b]
\tabletypesize{\scriptsize}
%\rotate
%\tablewidth{0pt}
%\tablenum{}
\tablecaption{Observations at 7 mm \label{obs7mm}}
%\tablehead{}
%\tablecolumns{}
\startdata
\hline \hline
                &              &   Bootstrapped    &                    &                           &                     \\ 
                &              &  Flux Density of  & \multicolumn{2}{c}{Synthesized Beam\tablenotemark{b}} &			  \\ \cline{4-5}
 Observation    &     VLA      & Phase Calibrator\tablenotemark{a}&	HPBW		& \multicolumn{1}{c}{P.A.}  & rms Noise\tablenotemark{b} \\
 Date\tablenotemark{a} &Configuration &      (Jy)	   &	(arcsec)	& \multicolumn{1}{c}{(deg)} & (mJy beam$^{-1}$)   \\ \hline
02-Jan-05       & D            & 2.38 $\pm$ 0.04   & 2.37$\times$1.75 & $-$22 & 0.12 \\
04-Feb-01       & CnB          & 1.85 $\pm$ 0.05   & 0.45$\times$0.39 & $-$76 & 0.12 \\
04-Mar-26       & C            & 2.22 $\pm$ 0.05   & 0.58$\times$0.51 & $-$8  & 0.12 \\ 
05-Jul-29       & C            & 1.68 $\pm$ 0.02   & 0.57$\times$0.51 & $+$34 & 0.12 
\enddata
\tablenotetext{a}{The phase calibrator used in all the observations was 1925+211.}
\tablenotetext{b}{For naturally weighted maps.}
\end{deluxetable*}

 Very Large Array (VLA) D-configuration observations (angular resolution of $\sim$8$\arcsec$) towards the center
of the outflow revealed two 3.6 cm continuum sources, VLA 1 and VLA 2, separated by 15$\arcsec$ (Anglada et al.
1991). However, only VLA 2 is associated with millimeter emission from circumstellar dust (Cabrit \& Andr\'e
1991), suggesting that it is the counterpart of IRAS 19156+1906. VLA 2 has also been associated with a nearby
($\sim$0$\farcs$7 northeast from VLA 2) H$_2$O maser as well as with NH$_3$ emission (Girart et al. 1997). Also,
only VLA 2 seems to be associated with the CO outflow as shown by interferometric observations with the Nobeyama
Millimeter Array (Hirano et al. 1998) and with the Berkeley Illinois Maryland Association (BIMA) array (Lee et al.
2002). In the 3.6 cm maps of Anglada et al. (1996), made with data from the A and B configurations of the VLA
(angular resolution of $\sim$0$\farcs$3), the centimeter emission of VLA 2 is elongated along a direction with a
P.A. $\simeq$ 118$^\circ$, that is close to that of the large pair of CO lobes. In addition, H$_2$ emission and
Herbig-Haro (HH) objects have been found in the region along a direction with a P.A. $\simeq$ 115$^\circ$, similar
to that of the elongation of VLA 2 (Vrba et al. 1986; Hodapp 1994; Palacios \& Eiroa 1999; L\'opez et al. 2006).
These results suggest that the outflow activity along the direction $\sim 100^\circ$-$118^\circ$ is driven by the
young protostellar object VLA 2. However, the identification of the exciting source of the smaller pair of CO
lobes at P.A.$\simeq$30$^\circ$ remained unclear. As already pointed out, one possibility is that VLA 2 is the
driving source of the two pairs of bipolar lobes, e.g., driving each pair in a different epoch as a consequence of
precession of the jet axis. The second possibility is the presence of another, still undetected source in the
vicinity of VLA 2, that could be driving the small pair of CO lobes.

 Other regions harboring multipolar outflows, when studied with
subarcsecond angular resolution (equivalent to physical scales of tens to
hundreds of AUs for the relatively nearby regions that are better studied)
and sensitivity at the level of tens of $\mu$Jy, have shown that the
multipolar outflow morphology is usually associated with the presence of
multiple YSOs at the center of the outflows: e.g., IRAS~16293$-$2422 (CO
outflow: Mizuno et al. 1990; radio continuum: Estalella et al. 1991);  
IRAS~20050+2720 (CO outflow: Bachiller et al. 1995;  radio continuum:
Anglada, Rodr\'{\i}guez \& Torrelles 1998); HH111 (CO outflow: Cernicharo
\& Reipurth 1996; radio continuum: Reipurth et al. 1999); L1634 (H$_2$
jet: Hodapp \& Ladd 1995; radio continuum: Beltr\'an et al. 2002);  HH288
(CO outflow: Gueth et al. 2001; radio continuum: Franco-Hern\'andez \&
Rodr\'{\i}guez 2003); OMC-1S (optical jets: Bally et al. 2000; radio
continuum: Zapata et al. 2004).

 In this paper we present sensitive, high angular resolution multi-epoch observations at 3.6 cm of the radio
sources detected in the central region of the L723 outflow. Our goals were to search for changes in the structure
of VLA 2 that could help us to clarify the origin of the quadrupolar morphology of the outflow and to search for
additional indications of multiplicity at very small scales. We also analyze VLA archive 7 mm continuum data of
the region.

\section{Observations}
\begin{deluxetable*}{cccccccc}
\tabletypesize{\scriptsize}
%\rotate
\tablewidth{0pt}
%\tablenum{}
\tablecaption{YSO Candidates Associated with VLA 2\label{tabla2}}
%\tablehead{}
%\tablecolumns{}
\startdata
\hline \hline
          & \multicolumn{3}{c}{Position (J2000)\tablenotemark{a}} & &\multicolumn{2}{c}{Flux Density (mJy)\tablenotemark{b}} & Spectral  \\ \cline{2-4} \cline{6-7}
Component &     RA        &   DEC         &  Error         & &      3.6 cm      &     7 mm           &   Index    \\  \hline
A         & 19 17 53.673  & +19 12 19.59  &  0$\farcs$01  & & 0.11 $\pm$ 0.04\tablenotemark{c}   &   0.7 $\pm$ 0.2    & +1.1 $\pm$ 0.3 \\
B         & 19 17 53.685  & +19 12 19.34  &  0$\farcs$01  & & 0.08 $\pm$ 0.02	& $<$0.37\tablenotemark{d}  &  $<$+0.9        \\
C         & 19 17 53.739  & +19 12 19.77  &  0$\farcs$07  & & 0.05 $\pm$ 0.02\tablenotemark{e} &  0.7 $\pm$ 0.2\tablenotemark{f} & +1.5 $\pm$ 0.4 \\
D         & 19 17 53.894  & +19 12 17.96  &  0$\farcs$09  & & $<$0.02\tablenotemark{d} &   0.6 $\pm$ 0.2    &  $>$+2.0       
\enddata
 \tablenotetext{a}{Positions for components A and B are derived from Gaussian fits 
to the sources in the 3.6 cm ``All-Epochs'' uniformly weighted map (Fig. \ref{fig3}a). Positions 
for components C and D are derived from the 7 mm map (Fig. \ref{fig4}). Units of right 
ascension are hours, minutes, and seconds, and units of declination are 
degrees, arcminutes, and arcseconds.}
 \tablenotetext{b}{Derived from Gaussian fits to ``All-Epochs'' naturally weighted maps.}
 \tablenotetext{c}{Uncertainty is estimated from the observed peak intensity variability in our multi-epoch observations
 (Table \ref{tabla3})}
 \tablenotetext{d}{4-$\sigma$ upper limit.}
 \tablenotetext{e}{Sum of the flux densities of VLA 2Ca and VLA 2Cb.}
 \tablenotetext{f}{Angularly resolved source. Deconvolved size is $\sim1\farcs2\pm0\farcs2$ at a P.A 
$\simeq90^\circ\pm7^\circ$.}
 \end{deluxetable*}

 The 3.6~cm continuum observations were carried out in six epochs, ranging from 1995 August 12 to 2000 December
16. The data from the 1995.6 observations were already reported by Anglada et al. (1996). The VLA of the National
Radio Astronomy Observatory (NRAO)\footnote{The NRAO is a facility of the National Science Foundation operated
under cooperative agreement by  Associated Universities, Inc.}\ was used in its A configuration, providing an
angular resolution of $\sim$0$\farcs$3. Absolute flux calibration was achieved by observing 3C~286, for which a
flux density of 5.20~Jy at 3.6~cm was adopted. Data editing and calibration were carried out using the
Astronomical Image Processing System (AIPS) package of NRAO, following the standard VLA procedures. The
observation dates, the bootstrapped flux densities of 1925+211, the phase calibrator used in all the observations,
as well as the parameters of the synthesized beams, and the rms noise of the naturally weighted maps are given in
Table \ref{tabla1}. Due to the weakness of the sources detected the signal-to-noise ratio per baseline was
insufficient to self-calibrate the data.

 For each epoch, we made cleaned maps with different values of the parameter ROBUST of IMAGR (Briggs 1995). The
ROBUST parameter ranges in value from +5 to $-$5, with the maps made with ROBUST=+5 (equivalent to natural
weighting of the visibility data) having the highest sensitivity but the lowest angular resolution and the maps
made with ROBUST=$-$5 (equivalent to uniform weighting) having the highest angular resolution but the lowest
sensitivity. For each value of the ROBUST parameter the map of each epoch was restored using a circular beam whose
HPBW is the geometric mean  of the major and minor axes of the synthesized beam obtained by concatenating the uv
data of all the six epochs (HPBW = 0$\farcs$29 for naturally weighted maps and HPBW = 0$\farcs$19 for uniformly
weighted maps). This average restoring beam was adopted to facilitate comparison between the different epochs and
to allow averaging of all the 3.6 cm images into a single image. We consider this procedure appropriate since the
major and minor axes of the synthesized beams of the individual observing sessions (see Table \ref{tabla1}) are
within $\sim$10\% of the adopted average HPBW.

 To improve the sensitivity and uv coverage, we have concatenated the uv data from all the epochs to obtain
single maps, which we will refer to as ``All-Epochs'' maps. Since some of the structures persist from one epoch
to another, while others change, the procedure of concatenating data could result in spurious effects in the
final images. Therefore, we checked that the maps obtained by concatenating all the uv data do not show
significant differences from the maps obtained by averaging all the single epoch maps. 

 The 7 mm continuum data were taken from the VLA archive. The data consist of four sets of observations in the D,
C and CnB configurations (see Table \ref{obs7mm}). Absolute flux calibration was achieved by observing 3C~286, and
a model image at 7 mm for this calibrator was used to calibrate the data in amplitude. In order to improve the
sensitivity of the final image, but keeping a balanced contribution from the different configurations, we have
concatenated all the available uv data giving the same weight to the D and C+CnB data. Since these data come from
different configurations, it is not possible to average the individual maps. From these uv data we obtained a 7 mm
``All-Epochs'' naturally weighted map.

\section{Results and Discussion}

\begin{figure*}
%\epsscale{0.80}
\plotone{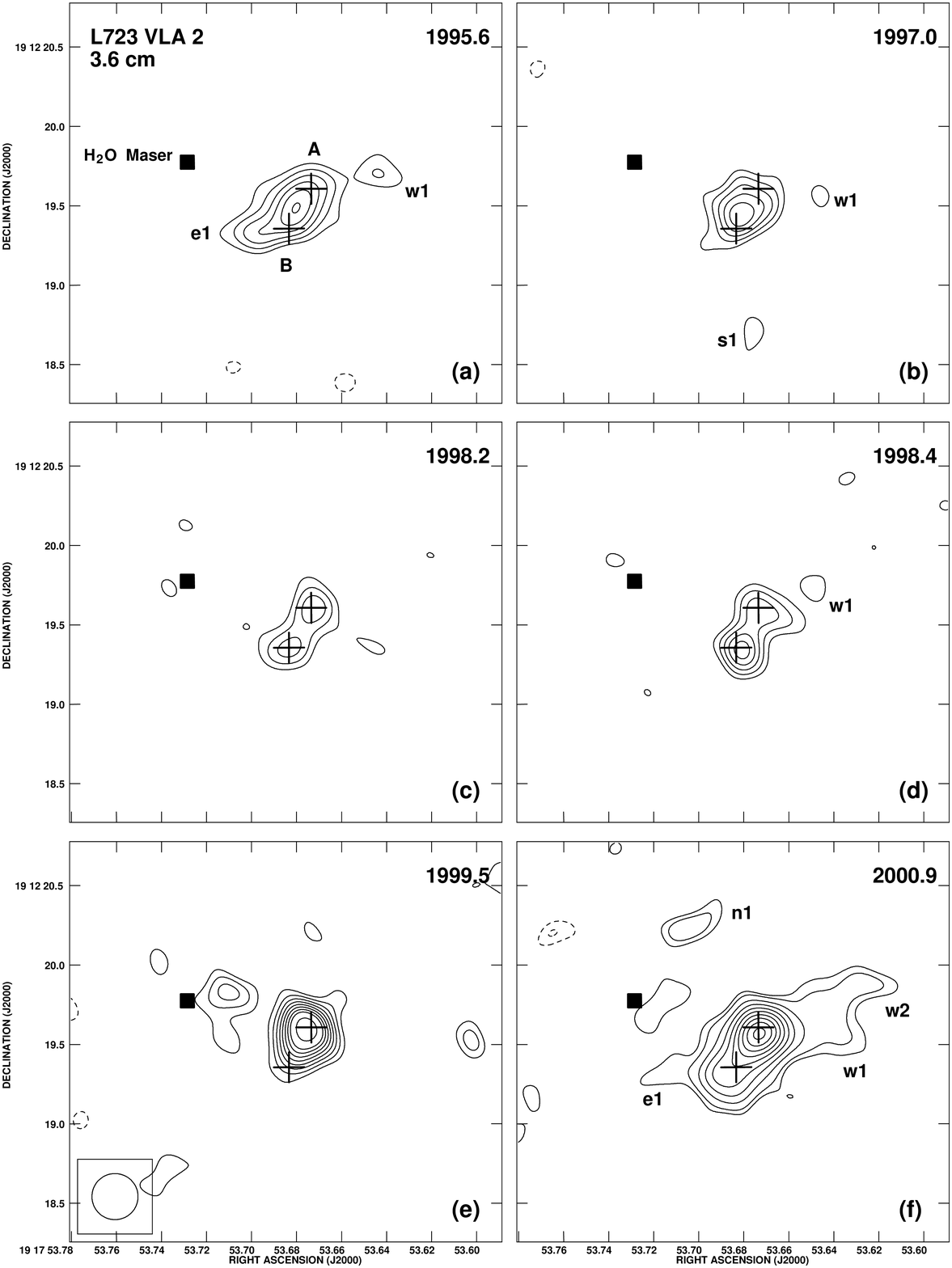}

\caption{\footnotesize{VLA 3.6 cm continuum maps of the VLA 2 region at different epochs, obtained with natural
weighting. In order to better identify morphological changes from one epoch to other, the maps have been restored
using the same circular beam of 0$\farcs$29 (HPBW) and contoured in terms of the same reference value (11 $\mu$Jy
beam$^{-1}$). For panels (a) to (e), contour levels are $-4$, $-$3, 3, 4, 5, 6, 7, 8, 9, 10, and 12 times 11
$\mu$Jy beam$^{-1}$, the typical rms of these maps (see Table \ref{tabla1}). For panel (f), additional contours at
$-$2, $-$1.5, 1.5, and 2 times 11 $\mu$Jy beam$^{-1}$ have been included so that the lowest contour level
corresponds approximately to 3 times the rms of this map (see Table \ref{tabla1}). The crosses mark the positions
of VLA 2A and VLA 2B derived from Gaussian fits to the uniformly weighted ``All-epochs'' map (see text). The filled
square marks the position of the H$_2$O maser detected by Girart et al. (1997). The radio continuum knots detected
in the region (w2, w1, e1, e2, s1, and n1) are also labeled (see \S 3.1).}}

\label{fig1}
\end{figure*}

\subsection{VLA 2A and 2B: A Close Binary System of YSOs}

 In Figure \ref{fig1} we show naturally weighted 3.6 cm maps of VLA 2 for each individual epoch. We have detected
radio continuum emission from VLA 2 in all the epochs, with total flux densities varying in the range 0.2-0.5 mJy.
In these maps, we see changes in the morphology of VLA 2, but some of the structures persist from one epoch to
another. In particular, in three of the epochs VLA 2 shows a clear double morphology with two components separated
by $\sim$0$\farcs$3 along a P.A. of $\sim$150$^\circ$. This double morphology appears in the 1998.2 and 1998.4
epochs, where the uv coverage was exactly the same, and in the 2000.9 epoch, which was a full-track observation
and therefore the uv coverage obtained was much better than in the other epochs (see Table \ref{tabla1} for the HA
ranges and time on source for each epoch of observation). In the 1999.5 epoch, the uv coverage was also very
similar to that of the 1998 epochs, but the double morphology is not so evident. We think this could be due to the
higher rms noise of the 1999.5 map and to the increase of the flux density of the northern component (see below)
that partially masks the emission of the southern component. Given these considerations, it is likely that VLA 2
is actually a persistent double radio source, with the differences in the observed morphology being mainly due to
the different uv coverage from one epoch to another.

\begin{deluxetable}{lccl}[!b]
\tabletypesize{\scriptsize}
%\rotate
\tablewidth{0pt}
%\tablenum{}
\tablecaption{Monitoring of the VLA 2 binary at 3.6 cm\label{tabla3}}
%\tablehead{}
%\tablecolumns{}
\startdata
\hline \hline
       &\multicolumn{2}{c}{Peak Intensity (mJy beam$^{-1}$)\tablenotemark{a}} & \multicolumn{1}{c}{Total Flux\tablenotemark{b}} \\ \cline{2-3}
Epoch	&   VLA 2A	   &   VLA 2B		& \multicolumn{1}{c}{Density (mJy)}    \\ \hline
1995.6  & 0.08 $\pm$ 0.01\tablenotemark{c}  &  0.06 $\pm$ 0.01\tablenotemark{c}	 &  0.30 $\pm$ 0.03    \\
1997.0  & 0.07 $\pm$ 0.01\tablenotemark{c}  &  0.06 $\pm$ 0.01\tablenotemark{c}	 &  0.18 $\pm$ 0.03    \\
1998.2  & 0.06 $\pm$ 0.01  &  0.06 $\pm$ 0.01   &  0.20 $\pm$ 0.02    \\
1998.4  & 0.06 $\pm$ 0.01  &  0.07 $\pm$ 0.01   &  0.16 $\pm$ 0.02    \\
1999.5  & 0.15 $\pm$ 0.01\tablenotemark{c}  &  0.05 $\pm$ 0.01\tablenotemark{c}	&  0.21 $\pm$ 0.03    \\
2000.9  & 0.13 $\pm$ 0.01  &  0.07 $\pm$ 0.01	&  0.24 $\pm$ 0.02    
\enddata
\tablenotetext{a}{Derived from Gaussian fits to the 3.6 cm naturally weighted maps that have been restored with an 
average circular beam of HPBW=0$\farcs$29 (Fig. \ref{fig1}).}
\tablenotetext{b}{Total flux density of the binary system excluding the extended emission (see Fig. \ref{fig1}).}
\tablenotetext{c}{Obtained by fixing the Gaussian centers to the positions of VLA 2A and VLA 2B given in Table \ref{tabla2}.}
\end{deluxetable}

\begin{figure}[!t]
%\epsscale{0.5}
\plotone{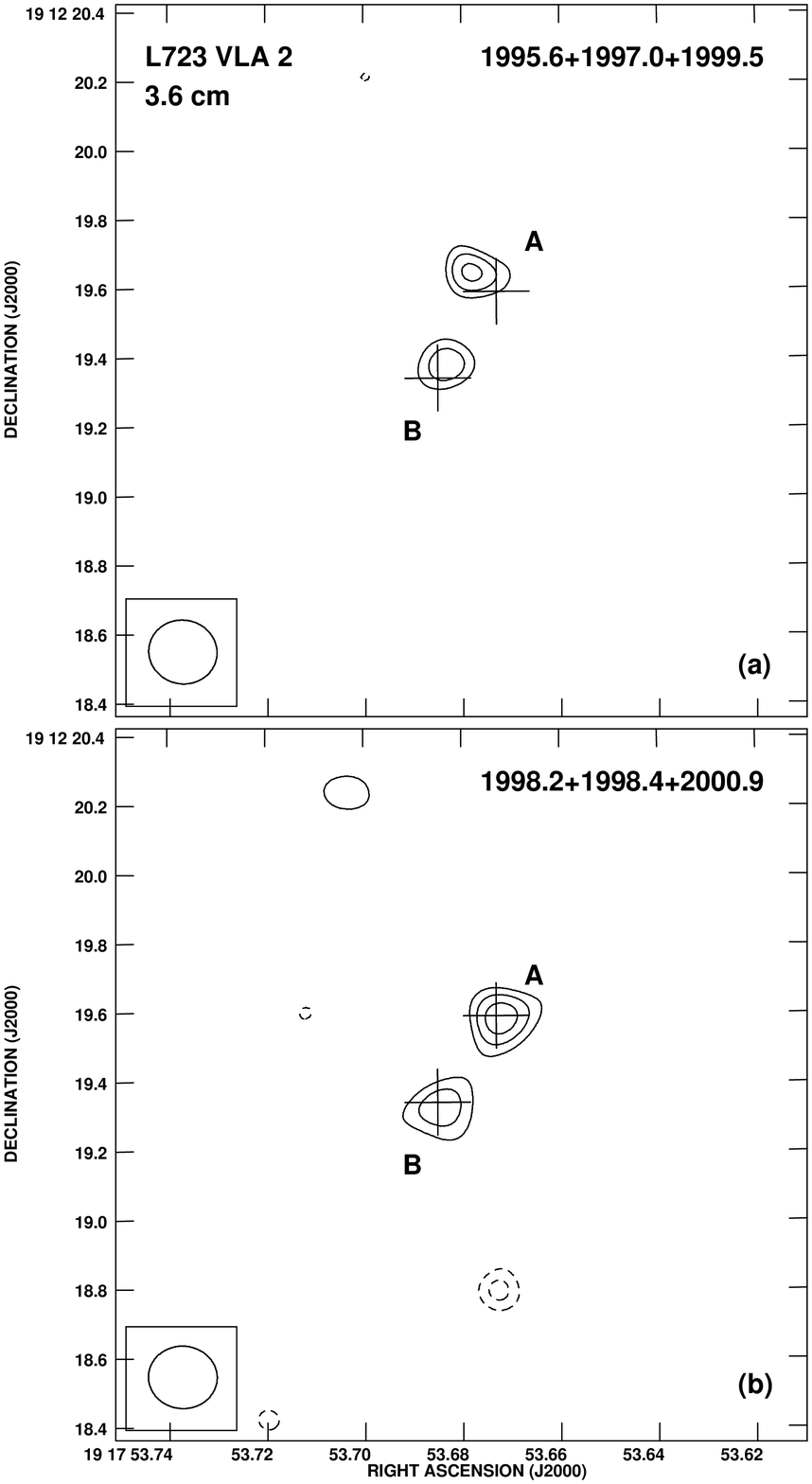}

\caption{\footnotesize{(a) Uniformly weighted map of the source VLA 2 at 3.6 cm made by concatenating the uv data of
the 1995.6, 1997.0, and 1999.5 epochs (where VLA 2 does not show a clear double morphology; see Fig. \ref{fig1}).
The synthesized beam of the map is 0$\farcs$20$\times$0$\farcs$18, with a P.A. of 79$^\circ$. Contour levels are
$-$4, $-$3, 3, 4, and 5 times the rms of the map, 17 $\mu$Jy beam$^{-1}$.(b) Same as (a), but for the 1998.2,
1998.4, and 2000.9 epochs (where VLA 2 shows a clear double morphology; see Fig. \ref{fig1}). The synthesized beam
of the map is 0$\farcs$20$\times$0$\farcs$18, with a P.A. of 86$^\circ$. Contour levels are $-$4, $-$3, 3, 4, and
5 times the rms of the map, 12 $\mu$Jy beam$^{-1}$. Note that in both maps the source VLA 2 shows a clear double
morphology, suggesting that VLA 2 is a binary system of embedded YSOs (see text for further details). The crosses
mark the positions of VLA 2A and VLA 2B derived from Gaussian fits to the uniformly weighted ``All-epochs'' map
(see text, and Fig. \ref{fig3}a).}}

\label{fig2}
\end{figure}

 A way to mitigate the effects introduced by the different uv coverages 
would be to make uniformly weighted maps of the individual epochs. In 
uniformly weighted maps all the grid cells have equal weight, which 
compensates for the high concentration of data points in the inner region 
of the uv plane. This provides a more uniform coverage of the gridded uv 
plane, resulting in maps having higher angular resolution than naturally 
weighted maps. Unfortunately, uniform weighting yields lower sensitivity, 
and due to the weakness of the source, it is difficult to detect it in the 
uniformly weighted map of a single epoch. However, it is still possible to 
test our hypothesis of a double radio source with the available data. For 
this purpose we have made uniformly weighted maps from two different sets 
of concatenated uv data: (i) the three epochs in which VLA 2 does not show 
clearly a double morphology, and (ii) the three epochs in which VLA 2 
shows a clear double morphology. This procedure allows us to compare two 
maps with high enough angular resolution and sensitivity, as well as a 
good uv coverage. As can be seen in Figure \ref{fig2}, VLA 2 shows always 
a double morphology (with a slight change of $\sim$0$\farcs$1 in the 
position of the sources), even in the map made with the data of the epochs 
in which the double morphology was not evident. This strongly supports 
that two radio continuum sources (that we call VLA 2A and VLA 2B) persist 
through all the epochs, covering a time span $\gtrsim$ 5 yr.

\begin{figure*}
\epsscale{0.9}
\plotone{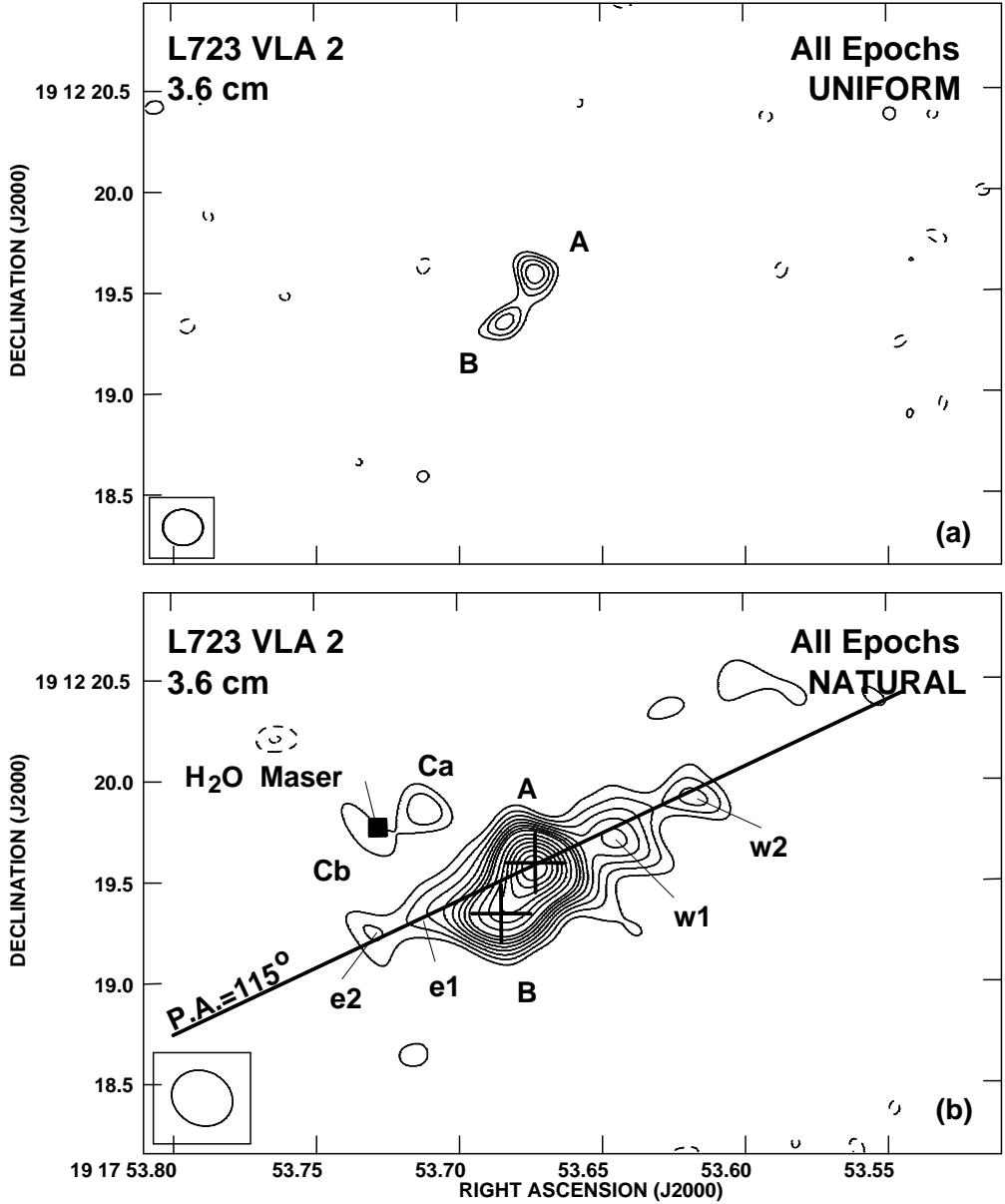}

\caption{\footnotesize{\emph{(a)} Uniformly weighted ``All-Epochs'' map of the source VLA 2 at 3.6 cm, obtained by
concatenating the uv data from all the epochs. The synthesized beam of the map is 0$\farcs$20$\times$0$\farcs$18,
with a P.A. of 86$^\circ$. Contour levels are $-$4, $-$3, 3, 4, 5, and 6 times the rms of the map, 11 $\mu$Jy
beam$^{-1}$. \emph{(b)} Same as (a), but with natural weighting. The synthesized beam of the map is
0$\farcs$31$\times$0$\farcs$27,  with a P.A. of 63$^\circ$. Contour levels are $-$4, $-$3, 3, 4, 5, 6, 7, 8, 9,
10, 12, 14, 16, 18, and 20 times the rms of the map, 4 $\mu$Jy beam$^{-1}$. Crosses mark the positions of VLA 2A
and VLA 2B derived from Gaussian fits to the uniformly weighted map. The filled square marks the position of the
H$_2$O maser detected by Girart et al. (1997). The axis of the extended emission (traced by knots w2, w1, e1, and
e2), at P.A. of $\sim$115$^\circ$, is indicated. }}

\label{fig3}
\end{figure*}

 Since all the data seem to be consistent with the presence of a double
source, we obtained an ``All-Epochs'' uniformly weighted map (Fig.  
\ref{fig3}a) by concatenating all the uv data. The positions and flux
densities of components VLA 2A and 2B, derived from Gaussian ellipsoid
fits to this map, are given in Table \ref{tabla2}. The angular separation
of the two components is 0$\farcs$29, which corresponds to a projected
separation of $\sim$90 AU (assuming a distance of 300 pc). If these
sources were tracing some kind of high-velocity outflowing features, such
as knots in a jet, a significant variation in their positions in the
single epoch maps would be expected (e.g., a velocity of $\sim$100 km
s$^{-1}$ in the plane of the sky would result in an angular displacement
of $\sim 0\farcs35$ over the period of $\sim$5 yr covered by our
observations). The lack of a significant displacement from one epoch to
another excludes that these sources trace high-velocity material, unless
the axis of the motion was very close the line of sight (which does not
seem to be the case, given that the blue and red lobes of the molecular
outflow appear clearly separated on the sky). This, together with the
additional properties observed for these sources (see below) led us to
interpret VLA 2A and VLA 2B as a close binary system of embedded YSOs.

  The individual epoch maps shown in Figure \ref{fig1} suggest variability in the 3.6 cm flux density of the
compact components of the double source. In  Table \ref{tabla3} we give the peak intensities of VLA 2A and VLA 2B
at each epoch of observation. Because the angular resolution of our observations is similar to the angular
separation of the two sources, these values of the peak intensity must be taken with care. However, given that the
maps of the different epochs have been restored with the same beam, the variations of the peak intensities are
expected to be representative of the variability of the individual components. The peak intensity of VLA 2B
appears to be constant at all epochs, consistent with a value of 0.06 $\pm$ 0.01 mJy  beam$^{-1}$ (see Table
\ref{tabla3}). On the other hand, the peak  intensity of VLA 2A is consistent with a value of 0.07 $\pm$ 0.01 mJy 
beam$^{-1}$ (see Table \ref{tabla3}) in the four first epochs, while in  the last two epochs the peak intensity of
the source increased approximately by a factor of two. Similar flux density variability has been  observed in
several embedded YSOs (e.g., Cep A, Porras et al. 2002; B335,  \'Avila et al. 2001). We interpret the variability
in the flux density of  VLA 2A as due to material recently ejected that still remains very close  to the YSO and
it is not distinguishable from the source.

\begin{figure*}
%\epsscale{1}
\plotone{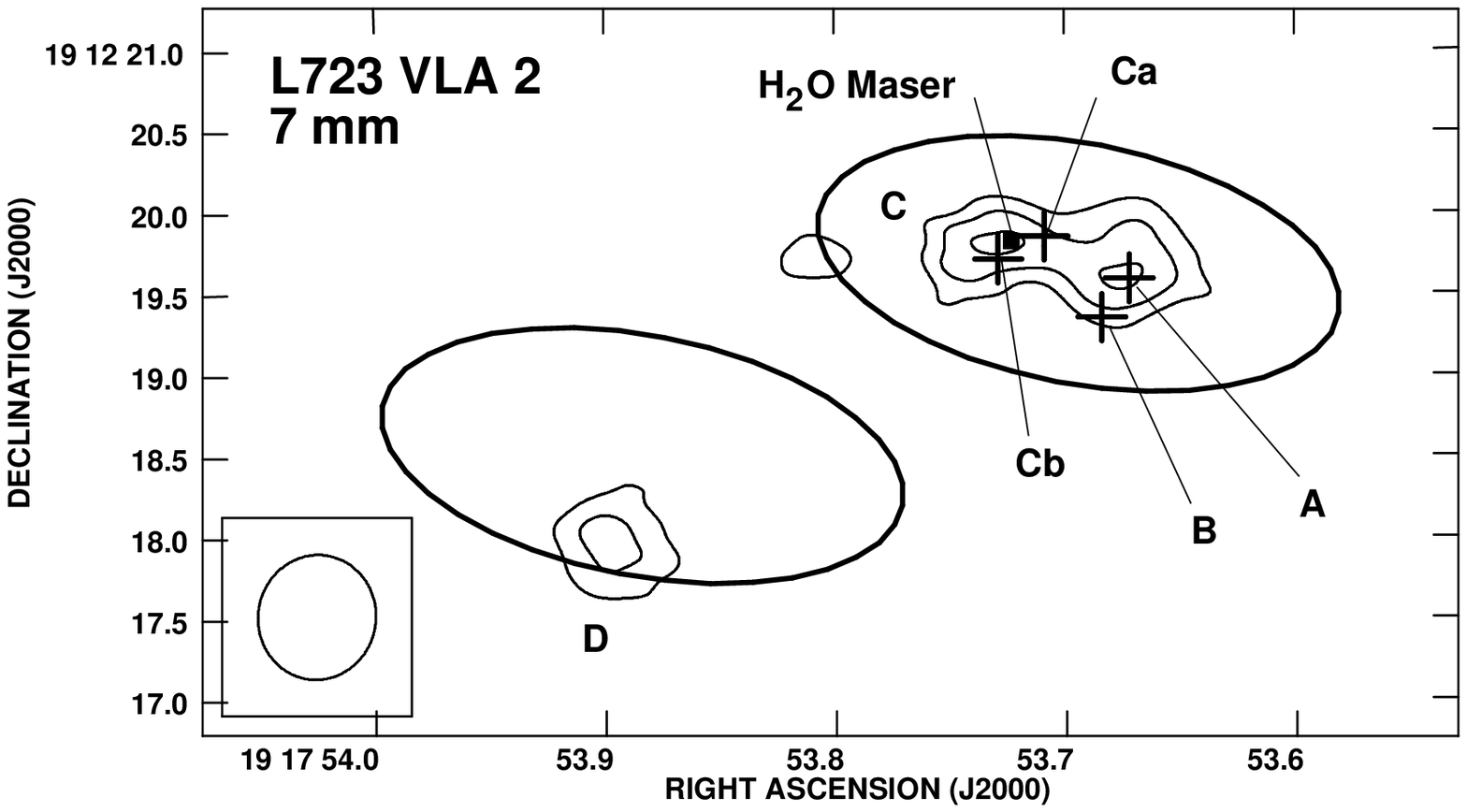}

\caption{\footnotesize{Naturally weighted map of the VLA 2 region at 7 mm, obtained by concatenating the uv data from VLA
D, CnB, and C  configurations. The synthesized beam of the map is 0$\farcs$77$\times$0$\farcs$73 with a P.A. of
$-10^\circ$. Contour levels are $-4$, $-$3, 3, 4, and 5 times the rms of the map, 93 $\mu$Jy beam$^{-1}$. Crosses
mark the positions of VLA 2A, VLA 2B, VLA 2Ca, and VLA 2Cb at 3.6 cm. The filled square marks the position of the 
water maser detected by Girart et al. (1997). Ellipses are centered at the  nominal positions of the 1.35 mm sources
detected by Girart et al. (2007), and their size corresponds to the synthesized beam of the SMA.}}

\label{fig4}
\end{figure*}

 In addition, we have detected 7 mm emission from VLA 2A (see Fig.
\ref{fig4}).  The flux densities at 3.6 cm and 7 mm of VLA 2A (see Table
\ref{tabla2}) imply a spectral index $\alpha$ $\simeq$ 1.1$\pm$0.3
(S$_\nu$ $\propto$ $\nu^\alpha$), consistent with partially optically
thick free-free emission that could be tracing a radio jet. The nominal
value of the spectral index obtained is slightly higher than the typical
values found in ionized jets driven by YSOs, 0.3 $\la$ $\alpha$ $\la$ 0.9
(Anglada 1996 and references therein).  This higher value of the spectral
index of VLA 2A could be indicating that we are partially missing the
extended emission of the possible radio jet, either because of
insufficient uv coverage or because part of this extended emission is
confused with the emission of the nearby source VLA 2B. Since the outer
part of a thermal radio jet is where the emission is optically thinner,
we would be observing preferentially the emission coming from the region
closer to the driving source, where the optical depth is higher. This
effect would increase the value of the spectral index with respect to that
of the overall emission of a radio jet. Alternatively, a high value of the
spectral index could be indicative of a significant contribution from dust
emission at 7 mm. Both possibilities (free-free emission from a jet with a
possible contribution of dust emission from a circumstellar disk or an
envelope) are consistent with VLA 2A being a YSO.

We should note that the spectral indices derived in this paper have been
obtained from non-simultaneous observations and can be affected by
time-variability of the sources.  However, as discussed above, the flux
variations observed in our multi-epoch observations suggest modest
variability that it is not expected to affect our conclusions on the
nature of the emission derived from the spectral index. Also, the 3.6 cm
and 7 mm observations have different uv coverages, that could introduce
additional errors in the flux density measurements of extended emission.
Finally, the presence of several sources in a small area makes difficult
to isolate the contribution of each source. Therefore, the spectral
indices reported in this paper should be considered as rough estimates.

 In addition to the compact emission of the two sources VLA 2A and VLA 2B,  in the individual 3.6 cm maps of
Figure \ref{fig1} we also note the presence of  several weak ``knots'' that could be tracing extended emission. In
order to gain sensitivity to extended structures, we have concatenated all the 3.6 cm uv data and made a naturally
weighted ``All-Epochs'' map, that is shown in  Figure \ref{fig3}b. This map shows extended emission with a P.A.
of  $\sim$115$^\circ$, where we identify the knots w1, w2, e1, and e2, that can also be seen in several of the
individual epoch maps (Fig.  \ref{fig1}). The extended emission seems to emanate from VLA 2A, the northern
component of the double source, since a line running through the  major axis of this extended emission is centered
close to the position of VLA 2A (see Fig. 3b), which is the strongest source along this direction.  The P.A. of
this extended emission ($\sim$115$^\circ$) coincides with that of the system of HH objects observed in the region
(L\'opez et al. 2006), suggesting that VLA 2A is a YSO that is driving these HH objects. The P.A. of the extended
emission is  also similar to that of the symmetry axis of the larger pair of CO outflow lobes ($\sim$100$^\circ$),
thus suggesting that VLA 2A is also associated with this lobe pair.

 We interpret the extended radio emission as tracing an ionized jet that originates at the position of VLA 2A, and
where the knots w1, w2, e1, and  e2 would be tracing electron density enhancements produced by  shock-ionization.
We do not detect clear proper motions associated with these knots from one epoch to another, setting an upper
limit to the proper motion velocity of the knots of $\sim$27 km s$^{-1}$ (however, due to the coarse temporal
sampling and the low signal-to-noise ratio of the knots in the individual epoch maps, they are difficult to
identify from one epoch to another, and we cannot discard higher proper motion velocities). This suggests that
these knots have been excited by the interaction of the jet of VLA 2A with dense, stationary clumps in the
surrounding gas.

 We also note the detection of the faint knots, n1 (in the 2000.9 image; Fig. \ref{fig1}f) and s1 (in the 1997.0
image; Fig. \ref{fig1}b). We interpret these two knots as shock-ionized gas associated with episodic ejections of
material. Their location, nearly aligned with source VLA 2B, at a P.A. $\simeq$ 20$^\circ$, is consistent with
being excited by a jet coming from this source. The P.A. of this jet is similar to that of the smaller pair of CO
outflow lobes ($\sim$30$^\circ$) and therefore, we tentatively propose VLA 2B as its exciting source.

\begin{deluxetable*}{cccccccc}
\tabletypesize{\scriptsize}
%\rotate
\tablewidth{0pt}
%\tablenum{}
\tablecaption{Monitoring of VLA 1 at 3.6 cm$^{\rm a}$\label{tabla4}}
%\tablehead{}
%\tablecolumns{}
\startdata
\hline \hline
        &              &              &             &                  & \multicolumn{3}{c}{Deconvolved Size} \\ \cline{6-8}
        &    \multicolumn{3}{c}{Position (J2000)\tablenotemark{b}}   &  Flux Density    &  Major Axis       &  Minor Axis     &  P.A.    \\ \cline{2-4}
Epoch	& RA           &       DEC    &    Error    &  (mJy)           & (arcsec)         &  (arcsec)       & (deg)    \\ \hline
1995.6  & 19 17 52.922 & +19 12 08.85 & 0$\farcs$01 &  0.36 $\pm$ 0.02 & 0.11  & 0.10 & 179  \\
1997.0  & 19 17 52.923 & +19 12 08.85 & 0$\farcs$01 &  0.38 $\pm$ 0.02 & 0.13  & 0.00 &  37  \\
1998.2  & 19 17 52.922 & +19 12 08.86 & 0$\farcs$01 &  0.34 $\pm$ 0.02 & 0.12  & 0.04 & 179  \\
1998.4  & 19 17 52.922 & +19 12 08.86 & 0$\farcs$01 &  0.31 $\pm$ 0.02 & 0.11  & 0.05 & 179  \\
1999.5  & 19 17 52.923 & +19 12 08.86 & 0$\farcs$01 &  0.34 $\pm$ 0.03 & 0.13  & 0.07 & 179  \\
2000.9  & 19 17 52.923 & +19 12 08.85 & 0$\farcs$01 &  0.34 $\pm$ 0.01 & 0.13  & 0.00 &  30  
\enddata	
\tablenotetext{a}{Parameters derived from Gaussian fits to the 3.6 cm naturally weighted maps that have been restored with an 
average circular beam of HPBW=0$\farcs$29.}
\tablenotetext{b}{Units of right ascension are hours, minutes, and seconds, and 
units of declination are degrees, arcminutes, and arcseconds.}
\end{deluxetable*}

In summary, we detected two compact 3.6 cm sources, VLA 2A and VLA 2B, 
separated by $\sim$0$\farcs$3 along a P.A. of $\sim$150$^\circ$. We also 
detected extended 3.6 cm emission, centered on VLA 2A, and along a P.A. of 
$\sim$115$^\circ$, that coincides with the P.A of the system of HH objects 
in the region and is similar to that of the larger pair of CO outflow 
lobes. The spectral index of VLA 2A in the 3.6 cm to 7 mm wavelength 
range, is consistent with this source tracing the origin of an ionized 
radio jet with a possible contribution from dust. Therefore, we conclude 
that VLA 2A is an embedded YSO associated with a radio jet; it is also the 
likely exciting source of the system of HH objects and of the larger pair 
of lobes of the CO outflow observed in the region. We identify several 
knots (w1, w2, e1, and e2) distributed along the extended 3.6 cm emission, 
at both sides of VLA 2A. Since these knots are much weaker than VLA 2A and 
are aligned along the axis of the extended emission we interpret them as 
tracing electron density enhancements in the jet from VLA 2A rather than 
tracing the location of other YSOs. The lack of clear proper motions 
suggests that these knots have been excited by the interaction of the jet 
with dense, stationary clumps in the gas surrounding VLA 2A. The compact 
source VLA 2B has a flux density comparable to that of VLA 2A and both 
sources are aligned along a P.A. of $\sim$150$^\circ$, a direction where 
there are no indications of outflow activity. Therefore, we interpret VLA 
2B as a second embedded YSO, constituting a close (projected separation of 
90 AU) binary system. Neither VLA 2A nor VLA 2B show noticeable proper 
motions; this result strongly suggests that these sources do not trace 
high-velocity material, and is consistent with these sources being YSOs. 
Unfortunately, with the current data it is not possible to obtain a 
precise enough measure of proper motions that would allows us to derive 
possible orbital motions, as has been done in other sources (e.g., Loinard 
et al. 2002); this kind of data could be helpful to confirm that this is a 
binary system. Also, there are no high angular resolution data outside the 
centimeter wavelength range that allow us to characterize the individual 
spectral energy distribution of each component in order to establish the 
evolutionary status of the sources and to confirm their nature as YSOs. 
Therefore, these sources are good targets for future observations with the 
EVLA and ALMA. We also detect in the single epoch images two faint knots 
(n1 and s1) nearly aligned with the source VLA 2B along a P.A. of 
$\sim$20$^\circ$, that is similar to the P.A. of the smaller pair of CO 
outflow lobes. We tentatively interpret these two knots as shock-ionized 
gas associated with episodic ejections from VLA 2B, that could be the 
driving source of these knots as well as of this pair of CO outflow lobes.

\subsection{Other Radio Sources Detected in the Region}

 In addition to VLA 2A, our 7 mm image (Fig. \ref{fig4}) shows emission from other two sources. One of them,
that we will refer to as VLA 2C, coincides with the position of the water maser emission detected by Girart  et
al. (1997) and Furuya et al. (2003). The coincidence of this source with the water maser suggests that VLA 2C is
a YSO. At 7 mm, the source VLA 2C appears elongated, with a deconvolved size of 1$\farcs$2 $\pm$ 0$\farcs$2 with
a P.A. $\simeq$ $90^\circ\pm 7^\circ$. At 3.6 cm, we detect weak extended emission that coincides with the
source detected at 7 mm. The 3.6 cm emission shows two peaks, that we call VLA 2Ca and VLA 2Cb. With the current
data it is unclear if these two peaks actually correspond to two objects or to a single extended source.
Sensitive high angular resolution multiwavelength observations would be required to further define the nature of
this emission. From the total flux density at 3.6 cm (VLA 2Ca $+$ VLA 2Cb) and 7 mm (see Table \ref{tabla2}), we
estimate a spectral index of 1.5$\pm$0.4 for VLA 2C, a value suggestive of thermal emission from a YSO (see
discussion in \S3.1).

 The third 7 mm source, VLA 2D, is located $\sim 3\farcs5$ from VLA 2A and  does not have a detected 3.6 cm
counterpart. However, this source is located within $\sim$0$\farcs$6 of the nominal position of a 1.35 mm  source
detected by Girart et al. (2007) with the Submillimeter Array (SMA) (see Fig. \ref{fig4}). Given that the beam size
of the SMA observations is 3$\farcs$3$\times$1$\farcs$5 (P.A.=79$^\circ$), we suggest  that VLA 2D is the 7 mm
counterpart of the 1.35 mm source. From the total flux density at 7 mm of VLA 2D, and adopting a 4$\sigma$ upper
limit to the 3.6 cm flux density, we estimate a lower limit to the spectral index of this source of $\sim$2.0 in this
wavelength range, suggesting that the 7 mm emission is dominated by thermal dust emission. Additionally, the 1.35 mm
source seems to be associated with an elongated SiO structure, probably tracing material shocked by a high velocity
flow (Girart et al. 2007), suggesting that this source is associated with a YSO.

 In addition to the sources VLA~2A, VLA 2B, VLA 2C, and VLA 2D that appear located at the center of the L723
outflow, our 3.6 cm images detect the source VLA 1, first reported by Anglada et al. (1991), and located
$\sim$15$\arcsec$ SW of the position of VLA 2A. The source VLA 1 appears always compact (deconvolved size
$\le0\farcs13$; see Table \ref{tabla4}) with a roughly constant flux density consistent at all epochs with a value
of 0.36$\pm$0.02 mJy (see Table \ref{tabla4}). We have not detected this source at 7 mm up to a 4$\sigma$ level of
0.32 mJy. With our 3.6 cm flux density and the 7 mm upper limit, we estimate for VLA 1 a spectral index
$\alpha\le-$0.1, a value consistent with the spectral index $\alpha$=$-$0.3$\pm$0.2 measured by Anglada et al.
(1996) between 3.6 and 6 cm. As discussed by Anglada et al. (1996) and Girart et al. (1997), this source is most
probably a radio-emitting, optically-obscured T Tauri star located very near to the denser part of the molecular
cloud. Our accurate astrometry indicates that this compact source does not show noticeable proper motions, with a
position displacement $< 0\farcs03$ (3$\sigma$) over the 5.3 yr period covered by our observations (see Table
\ref{tabla4}). Since a total proper motion displacement of $\sim0\rlap.{''}03$ is expected for an object embedded
in the L723 cloud (assuming a distance of 300 pc and the Brand \& Blitz 1993 model of galactic rotation), this
result is compatible with VLA 1 belonging to the L723 cloud. However, the possibility that VLA 1 is a
line-of-sight source not directly associated with the molecular cloud is not discarded. The remaining radio
sources (VLA 2A, VLA 2B, VLA 2C, and VLA 2D) discussed in this paper are embedded in regions of extended radio
emission and are associated with ejecta, which does not allow a precise determination of their possible proper
motions.

\subsection{The Morphology of the Outflow in L723: Three Bipolar Outflows?}

\begin{figure*}  
%\epsscale{0.80}  
\plotone{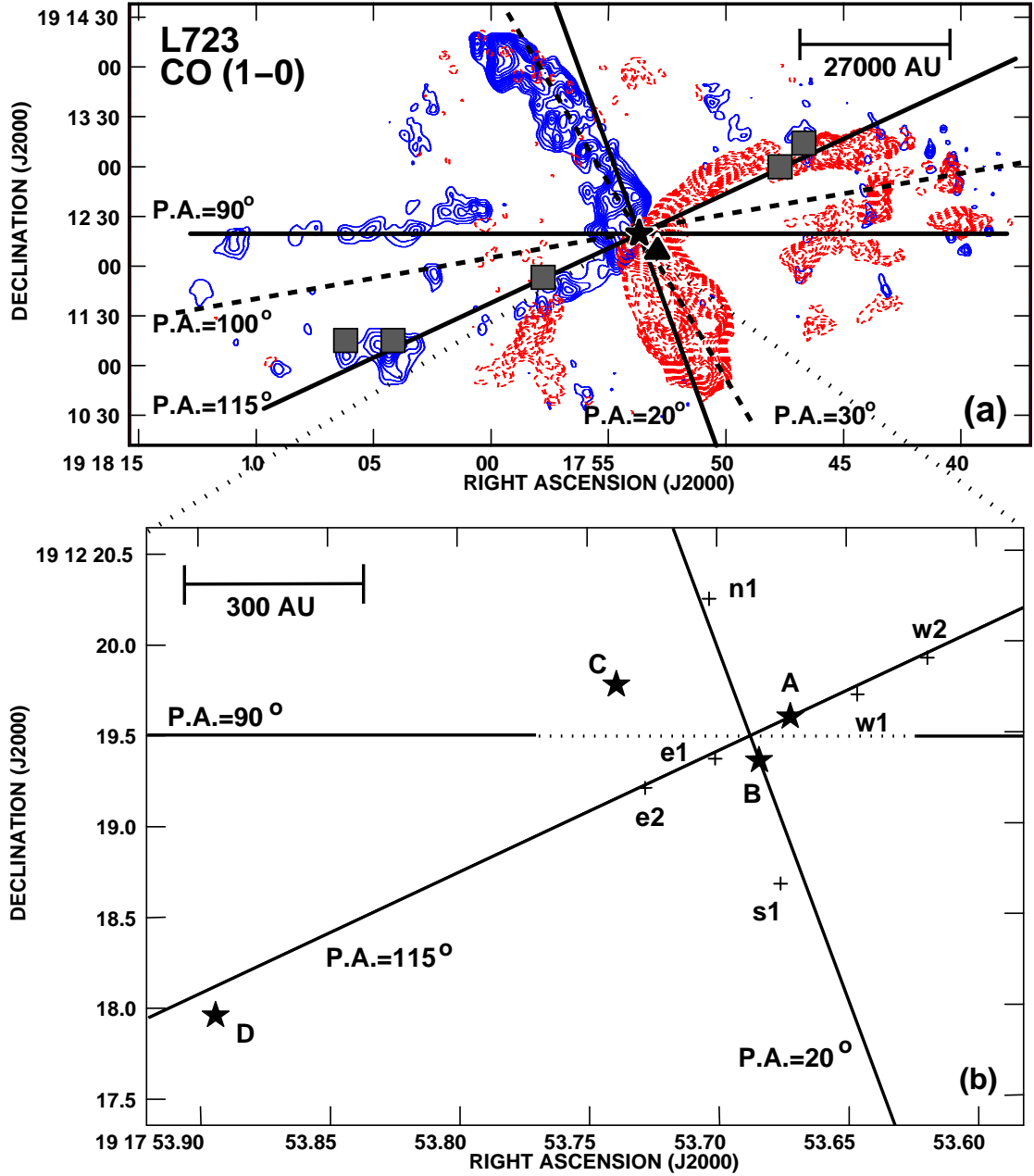}  

\caption{\footnotesize{ \emph{(a)} Map of the high velocity CO  ($J$=1$\rightarrow$0) emission in L723 (Lee et al.
2002). Redshifted  emission (dashed contours) is integrated from 11 to 31 km s$^{-1}$ and  blueshifted emission
(solid contours) is integrated from $-$10 to 10 km  s$^{-1}$. Dashed lines mark the position angles of the axes of
the bipolar outflows proposed by Lee et al. (2002). Solid lines mark the position  angles of the axes of the three
bipolar outflows proposed in this paper. Herbig-Haro objects (L\'opez et al. 2006) are marked by boxes. The star 
symbol marks the position of VLA 2 and the triangle marks the position of  VLA 1 (Anglada et al. 1991, 1996).
\emph{(b)} A schematic depiction of the  proposed outflow axes and the multiple system of YSOs at the center of the 
L723 outflow. Star symbols mark the positions of the suspected YSOs: VLA 2A, VLA 2B, VLA 2C, and VLA 2D. The three
axes have the directions of the outflows proposed in this paper. The crosses mark the positions of the knots detected
at 3.6 cm.}}

\label{fig5}  
\end{figure*}

In the maps of Avery et al. (1990), the L723 CO outflow shows a four-lobe morphology with a large pair of bipolar
lobes along a P.A. of $\sim 100^\circ$ and a smaller pair of bipolar lobes along a P.A. of $\sim 30^\circ$. As
discussed in the previous sections, our radio continuum data suggest that a multiple system of at least four YSOs
is present, within a region of $\sim 1200$ AU, at the intersection of the axes of symmetry of the two pairs of CO
lobes. Furthermore, in \S 3.1 we discussed the results that led us to propose that VLA 2A is related to the
excitation of the larger pair of CO lobes, while VLA 2B could be driving the smaller pair of CO lobes.

 Nevertheless, the interferometric maps of Lee et al. (2002) (obtained with higher angular resolution than the maps
of Avery et al. 1990) reveal additional details of the CO outflow structure (see Fig.~\ref{fig5}a). As can be seen in
this figure, the smaller pair of CO lobes shows a compact shape with a well defined axis at P.A. $\simeq 30^\circ$.
On the other hand, the shape of the larger pair of CO lobes appears clumpy, with only a few weak CO emission knots
near the symmetry axis of this lobe pair (at P.A. $\simeq 100^\circ$), and most of the emission concentrated at both
sides of this central axis, roughly at P.A. $\simeq 90^\circ$ and P.A. $\simeq 115^\circ$. The distribution of the CO
emission of this larger lobe pair was interpreted by Lee et al. (2002) as tracing the limb-brightened cavity walls of
a bipolar outflow with a P.A. $\simeq 100^\circ$. However, we note that the distribution of the CO emission at both
sides of the central axis (P.A. $\simeq 100^\circ$) of this lobe pair is asymmetric. The emission at P.A. $\simeq
115^\circ$ starts at the center of the outflow (the position of VLA 2) and is stronger than the emission at P.A.
$\simeq 90^\circ$, which shows a gap, with the bulk of the emission apparently starting $\sim 1'$ away from the
outflow center. Furthermore, the P.A. of $115^\circ$ coincides with that of the extended radio emission associated
with VLA 2A and with that of the system of HH objects.

 These considerations led us to propose a new interpretation of the high-velocity CO emission in L723 in terms of
three bipolar outflows. One bipolar outflow at P.A. $\simeq 30^\circ$, that could be driven by VLA 2B. A second
bipolar outflow at P.A. $\simeq 115^\circ$, which is associated with a system of HH objects and shock-ionized
extended radio continuum emission with the same position angle, that is driven by VLA 2A. Finally, a third bipolar
outflow at P.A. $\simeq 90^\circ$, whose exciting source is uncertain. The lobes of the proposed outflows at P.A.
$\simeq 30^\circ$  and P.A. $\simeq 115^\circ$ appear well ``connected'' with the outflow center, suggesting recent
outflow activity in these directions. On the contrary, the lobes of the outflow at P.A. $\simeq 90^\circ$ are weaker
and well detached from the central region, suggesting that they resulted from activity taking place in the past.
Therefore, we speculate that this outflow at P.A. $\simeq$ 90$^\circ$ could be a ``fossil'' outflow in the sense that
its exciting source has not been active in the recent past. The motion observed will then be the result of momentum
conservation of the gas accelerated when the exciting source was active. A similar fossil  molecular outflow has been
proposed to exist in the molecular cloud B5 (Yu, Billawala, \& Bally 1999).

\section{Conclusions}

 Our high angular resolution 3.6 cm and 7 mm continuum observations towards the previously known source VLA 2, at
the center of the L723 multipolar CO outflow, have revealed a group of at least four YSOs within a region of
$\sim 4''$ (1200  AU) in size. The size of this region is comparable with the size of the region where Girart et
al. (1997) found a local heating of the high density gas. Two of our continuum sources, VLA 2A and VLA 2B, form a
close radio binary whose components are separated by $\sim$0$\farcs$29 ($\sim$ 90~AU). Another source, VLA 2C, is
associated with a water maser, and a fourth source, VLA 2D, that we  only detect at 7 mm, appears to be the
counterpart of a recently discovered 1.35 mm source.

 We propose that the multipolar CO outflow in L723 could result from the superposition of at least three
independent bipolar outflows with position angles of $\sim$115$^\circ$, $\sim$90$^\circ$, and $\sim$30$^\circ$,
driven by three different YSOs. Our 3.6 cm observations suggest that VLA 2A is  associated with an ionized jet at
a position angle of $\sim$115$^{\circ}$. We propose that this is the exciting source of the system of HH objects
previously detected in the region, as well as of a bipolar CO outflow with the same position angle. The presence
of two 3.6 cm knots aligned with VLA 2B at a  P.A. of $\sim$20$^\circ$ suggests that VLA 2B could be the exciting
source of the bipolar CO outflow at a P.A of $\sim$30$^\circ$. Finally, the third bipolar CO outflow, with a
position angle of $\sim$ 90$^{\circ}$, seems to be a  ``fossil'' outflow whose exciting source has not been very
active in the recent past.

\emph{Acknowledgements.} C.C.-G. acknowledges support from MEC (Spain) FPU fellowship. G.A.,  C.C.-G., J.M.G.,
M.O. and J.M.T. acknowledge support from MEC (Spain)  grant AYA 2005-08523-C03 (including FEDER funds). G.A.,
C.C.-G. and M.O. acknowledge partial support from Junta de Andaluc\'{\i}a (Spain). L.F.R. acknowledges the
support of DGAPA, UNAM, and of CONACyT (M\'exico). We acknowledge Chin-Fei Lee for providing us the FITS file of
the CO emission  in L723.

\newpage

\end{document}